\documentclass[11pt]{article}

\usepackage[english]{babel}

\usepackage{graphics,color}
\usepackage{newlfont,epsfig,amssymb,amsmath}

\begin{document}
	\date{}
	\begin{center}
		{\Large\bf Quantum phase estimation using squeezed quasi-Bell states}
	\end{center}
	\begin{center}
		{\normalsize Douglas Delgado de Souza and A. Vidiella-Barranco \footnote{vidiella@ifi.unicamp.br}}
	\end{center}
	\begin{center}
		{\normalsize{Gleb Wataghin Institute of Physics - University of Campinas}}\\
		{\normalsize{ 13083-859   Campinas,  SP,  Brazil}}\\
	\end{center}
%
\begin{abstract}
In this paper we present a study of the quantum phase estimation 
problem employing continuous-variable, entangled squeezed coherent (quasi-Bell)
states as probe states. We show that their inherent 
squeezing and entanglement properties might bring advantages, increasing the 
precision of phase estimation compared to protocols which employ 
other continuous variable states e.g., two-mode, entangled coherent states or 
single-mode, squeezed states. We also analyze the phase estimation 
process considering: i) a linear (unitary) perturbation, and ii) dissipation,  
and conclude that the use of entangled squeezed coherent states as probe states
may still be advantageous even under non-ideal conditions.
\end{abstract}
\section{Introduction}
The precision of measurements is a matter of utmost importance in various fields. It is fundamental not only for the 
development of engineering in general, but also in experiments that require extremely high precision, such as the 
detection of gravitational waves \cite{ligo16}, atomic clocks synchronization \cite{holland14}, precision thermometry 
\cite{sanpera15}, and magnetometry \cite{retzker15}. Even though the laws of quantum mechanics impose limits on the accuracy 
of measurements, this unwanted quantum noise can be circumvented by using special quantum effects, such as squeezing
and/or entanglement \cite{maccone04,maccone11}. For instance, one of the first proposals for quantum noise reduction in
interferometry was based on the mixing, in a beam-splitter, of a squeezed vacuum state (in place of ordinary vacuum) with a 
coherent state \cite{caves81}. In the past years then, we have been witnessing the establishment of the field of quantum 
parameter estimation. Suppose one wants to estimate an unknown parameter 
$\phi$ by quantum means. As a first step we should prepare a quantum state $|\Psi\rangle$ (probe state) of a suitable physical system. 
Subsequently, the parameter $\phi$ is somehow encoded in the quantum state via a unitary transformation applied to the probe 
state. After performing suitable measurements in the modified state, the corresponding experimental results can be used to 
estimate the parameter $\phi$. Clearly, the precision in estimating a parameter will depend on the quantum probe 
state employed in the process. As a matter of fact, non-classical states of light have become a promising resource 
for the improvement of measurement techniques \cite{burnett93,huelga97,takeuchi07,furusawa12}. As mentioned above, 
the first quantum states to be recognized as being useful for metrology, e.g., for phase estimation, were the single mode squeezed 
states \cite{caves81}, followed by the archetypal (entangled) N$00$N states \cite{dowling00}. However, the performance 
of parameter estimation based on N$00$N states is severely hampered by losses, as they are particularly fragile against dissipation. 
Yet, after a while it was found that N$00$N-type states based on coherent states of light (rather than Fock states) could 
allow a superior performance for phase estimation in the presence of small losses \cite{gerry00,gerry02,joo11,joo12}. Also, 
single-mode squeezed states, are not only convenient for phase estimation \cite{monras06,douglas14}, but their use 
as probe states may bring advantages even under non-ideal conditions \cite{douglas14}. Furthermore, we find in the literature 
several proposals aiming to improve the estimation process employing other types of probe states. For instance, there are schemes 
based on entangled Fock states \cite{dowling08} and cluster states \cite{jaksch09}. We also find proposals using  
amplified Bell states \cite{james13}, ``cat states" \cite{lee15}, 
entangled states with coherent states \cite{davidovich16,kim19} as well as general path-symmetric entangled states \cite{nha16}. 
An alternative method using a different type of interferometer [SU(1,1)] is discussed in reference \cite{zhang20}.
At the same time, the spectrum of possible applications of quantum phase estimation has widened, and among the recent proposals 
we may cite the solution of generalized eigenvalue problems \cite{parker20}, an atomic gyroscope \cite{wang20}, and distributed
quantum phase estimation, relevant for sensor networks with spatially distributed parameters \cite{pan21}. We remark that
some of the above mentioned works are based on states which do not have a N$00$N states' structure. 
Yet, although it is apparent that squeezing and entanglement are resources that may bring advantages to phase 
estimation, in general those effects have been considered separately in the existing protocols. One of the exceptions is 
probably reference \cite{james13}, where a kind of squeezed number (entangled) state is employed. Therefore, in our view,
further investigation is needed in order to better understand the roles of squeezing and entanglement in quantum 
estimation process. 

In this work we will discuss, for phase estimation purposes, the possibility of using as probe continuous-variable 
entangled states composed by squeezed coherent states $|\alpha,\xi\rangle$, the so-called 
quasi-Bell states \cite{hirota}. Our aim is to investigate in what extent squeezing and entanglement, non-classical 
features present simultaneously in the above mentioned states, can be relevant in the process of quantum aided metrology. 
In order to do so, we employ the quantum Fisher information (QFI), here denoted as $H$, a well known figure of 
merit used to quantify the precision in parameter estimation for a given probe state. The QFI
is a quantum generalization of the Fisher information, being obtained via a maximization over all possible 
measurement strategies \cite{helstrom76,paris09}, i.e., it does not depend on a specific measurement scheme.
A parameter $\phi$ can be estimated from the experimental results via a statistical estimator,
here denoted as $\hat{\phi}$. The achievable precision in the estimation of a parameter $\phi$ 
is constrained by the QFI according to the quantum Cram\'er-Rao inequality \cite{paris09}
\begin{equation}
    \delta\hat{\phi} \geq \frac{1}{H(\phi)},
\end{equation}
where $\delta\hat{\phi} = \mathbb{E}\left[ \left( \hat{\phi} - \phi\right)^2 \right]$\footnote{Here 
$\mathbb{E}\left[\hat{\phi}\right] = \int \hat{\phi}(x) p(x,\phi)dx$ is the expectation value of the
estimator $\hat{\phi}$, with $p(x,\phi)$ being the probability of observing $x$ given the parameter 
$\phi$.} is the variance of the estimator $\hat{\phi}$. This means that, the larger the QFI, the
larger will be the precision in the estimation process. In addition to the fact that the QFI is a useful 
and well-established tool for parameter estimation, it also allows straightforward comparisons with previous 
findings, e.g., estimation schemes using squeezed, non-entangled states 
\cite{monras06,douglas14}, as well as other kinds of entangled probe states.
We would like to remark that if one takes into account external unwanted influences such as noise 
\cite{huelga97} or even a ``unitary disturbance" \cite{douglas14,pasquale13}, the accuracy of the estimation may be 
considerably degraded. Hence, we will analyze the phase estimation using squeezed quasi-Bell states taking into account the 
effect of a linear unitary disturbance in the system. Besides, we will also discuss the destructive action of losses in the 
estimation process using quasi-Bell states, by computing an upper bound for the QFI in the presence of a dissipative 
environment. Our manuscript is organized as follows: 
in Section 2 we present the calculations of the QFI and study the quantum phase estimation using squeezed quasi-Bell states, 
and discuss the roles of entanglement and squeezing in the ideal case. Subsequently, we investigate the effect of a linear 
disturbance and also calculate an upper bound for the QFI in the case of having losses. In Section 3 we present our 
conclusions.

\section{Quantum phase estimation with squeezed quasi-Bell states}

A ``squeezed" version of quasi-Bell states may be defined as
\begin{eqnarray}
	|\Psi_{l}\rangle&=&{\cal {N}}\left(\left|\alpha,\xi\right\rangle _{A}\left|-\alpha,\xi\right\rangle _{B}
	+l\left|-\alpha,\xi\right\rangle _{A}\left|\alpha,\xi\right\rangle _{B}\right)\label{quasibell}\\ \nonumber 
	\\ \nonumber
	|\Phi_{l}\rangle&=&{\cal {N}}\left(\left|\alpha,\xi\right\rangle _{A}\left|\alpha,\xi\right\rangle _{B}
	+l\left|-\alpha,\xi\right\rangle _{A}\left|-\alpha,\xi\right\rangle _{B}\right)\,.
\end{eqnarray}
where $\xi=r e^{i\theta}$ is the squeezing parameter, with $\alpha,r\ge 0$
and ${\cal {N}}$ is the normalization factor ${\cal {N}}=1/\sqrt{1+l\left(l+2\kappa^{2}\right)}$. The component
states, i.e., the single mode squeezed coherent states are defined as $|\alpha,\xi\rangle = D(\alpha) {S}(\xi) |0\rangle$, 
being ${S}(\xi) = \exp(\xi^*\hat{a}^2-\xi\hat{a}^\dagger{}^2)$ the squeezing operator and 
$D(\alpha) = \exp\left(\alpha\hat{a}^\dagger - \alpha^* \hat{a}\right)$ Glauber's
displacement operator, the generator of a coherent state $|\alpha\rangle$ from the vacuum state; 
$|\alpha\rangle = D(\alpha) |0\rangle$.
The overlap between the different component states is given by
\begin{equation}
    \kappa=\langle{\alpha,\xi}|{-\alpha,\xi}\rangle=\exp\left\{-2\alpha^{2}\left[\cosh(2r)+\sinh(2r)\cos\theta\right]\right\}.
\end{equation}
Here we introduce the auxiliary interpolating parameter $l$ $(-1\le l\le1)$, which allows us to verify
the role of entanglement in a straightforward way. 
The states $|\Psi_{l}\rangle$ and $|\Phi_{l}\rangle$ have basically the same entanglement properties, and thus
we will consider only the input states $|\Psi_{l}\rangle$ here. 
We remark that the continuous variable entangled states above are 
particularly suitable for the analysis of the influence of both entanglement and squeezing via a procedure 
similar to the one adopted in reference \cite{douglas14}, where the single mode states $|\alpha,\xi\rangle$ are
employed as probe states.

Normally the generation of non-classical states is not an easy task, and one should be concerned 
about how the probe states could be created. We find in the recent literature a proposal \cite{tavassoly17}
for generation of entangled squeezed coherent states having either $l=1$ or $l=-1$, which are of the type 
here employed. The scheme is based in a cavity QED setup, consisting in a conducting cavity driven by a
classical field. We point out that the entangled squeezed coherent states can also be written as 
(e.g., for $l=1$)
\begin{equation}
	|\Psi\rangle = \hat{S}_A(\xi)\hat{S}_B(\xi)
	\left(\left|\alpha'\right\rangle _{A}\left|-\alpha'\right\rangle _{B}+\left|-\alpha'\right\rangle _{A}\left|\alpha'\right\rangle _{B}\right).
\end{equation} 
In other words, the state $|\Psi\rangle$ is equivalent to an entangled 
coherent state \cite{sanders92} under the action of local operations (single-mode squeezing). 
The interpolation parameter $l$, which is intimately related to the entanglement of the state 
$|\Psi_l\rangle$, will allow us to identify the role of entanglement in the phase estimation process without 
compromising the other parameters involved. For instance, we can set a value to the squeezing parameter and see how the QFI depends on the
amount of entanglement. The entanglement in the quasi-Bell state $|\Psi_l\rangle$ can be quantified by the Entanglement entropy $E(\rho)$,
which can be written as a function of the concurrence function $C$ as \cite{wootters97,wootters98} 
\begin{equation}
    E(\rho) = {\cal E}(C) = {\cal H}\left( \frac{1 + \sqrt{1 - C^2}}{2}\right),
\end{equation}
where 
\begin{equation}
    {\cal H}(x) = -x\log_2 x - (1-x)\log_2 (1-x).
\end{equation}
The concurrence of the states $|\Psi_l\rangle$ and $|\Phi_l\rangle$ is the same, that is
\begin{equation}
    C\left(|\Psi_l\rangle\right) = C\left(|\Phi_l\rangle\right) = \frac{2(1 - \kappa^2)|l|}{l^2 + 2 \kappa^2 l + 1}.
\end{equation}
We have that, for $l = -1$ the state $|\Psi_l\rangle$ has the same amount of
entanglement as a maximally entangled pair of qubits, while for $l = 1$ the state is partially entangled in general, and attains the maximum 
entanglement only in the limit of $\kappa\rightarrow 0$. For $l=0$ we have a product state, and mode $A$ of $|\Psi_l\rangle$ is reduced to the 
squeezed coherent state $|\alpha,\xi\rangle$. We recall that partially entangled states may be useful for implementing quantum information 
tasks \cite{rigolin10,rigolin13}.

\subsection{\label{subsec1}Phase estimation without disturbance}

\begin{figure}
	\begin{centering}
		\includegraphics[height=3cm]{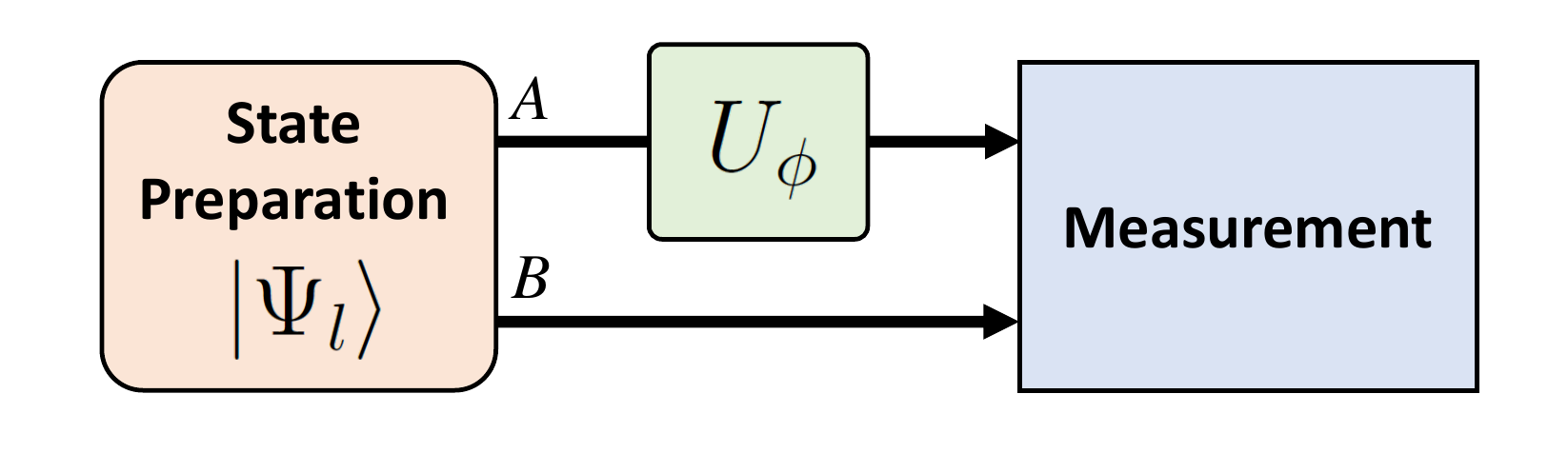}
		\caption{\label{setup} Schematic illustration of the setup for estimation of parameter $\phi$: i) a quasi-Bell, two-mode entangled state 
			$|\Psi_l\rangle$ is prepared beforehand; ii) one of the modes (mode $A$) is submitted to a phase shift 
			$U_{\phi}=\exp\left(-i\phi \hat{a}_{A}^{\dagger}\hat{a}_{A}\right)$, where the unknown phase $\phi$ is to be
			estimated; iii) the system, now in the modified state, undergoes a measurement.}
	\end{centering}
\end{figure}

We benchmark the phase estimation performance of the quasi-Bell states in a setup as shown in Fig. \ref{setup}.
Having prepared the probe states $|\Psi_{l}\rangle$, the parameter $\phi$ is encoded in mode $A$ of state 
$|\Psi_l\rangle$ via a transformation (phase shift) performed before the measurement procedure, say 
$|\Psi'_l(\phi)\rangle = U_{\phi}|\Psi_l\rangle$, where 
\begin{equation}
	U_{\phi}=e^{-i\phi G_{A}}, \ \ \ \ \ G_{A}=\hat{a}_{A}^{\dagger}\hat{a}_{A}\,.
\end{equation}

Since the transformation above is unitary and we are dealing with pure states, the QFI is simply given by \cite{monras06,douglas14}:
\begin{equation}
	H\left(\phi\right) = 4\left[\langle\Psi_{l}|G_{A}^{2}|\Psi_{l}\rangle-\left(\langle\Psi_{l}|G_{A}|\Psi_{l}\rangle\right)^{2}\right].\label{qfimain}
\end{equation}
Expanding the expression in Eq. (\ref{qfimain}) we obtain:
\begin{eqnarray}
	H\left(\phi\right) & = & 4{\cal{N}}^2\Big(_{A}\langle{\alpha,\xi}|G_{A}^{2}|{\alpha,\xi}\rangle_{A}+
	l^{2}{}_{A}\langle{-\alpha,\xi}|G_{A}^{2}|{-\alpha,\xi}\rangle_{A}+\\
	&  & +l\kappa{}_{A}\langle{-\alpha,\xi}|G_{A}^{2}|{\alpha,\xi}\rangle_{A}+
	l\kappa{}_{A}\langle{\alpha,\xi}|G_{A}^{2}|{-\alpha,\xi}\rangle_{A}\Big)-\left(n_{\mathsf{in}}\right)^{2},
\end{eqnarray}
where $n_{\mathsf{in}}$ is the mean photon number of the state $|\Psi_l\rangle$,
\begin{equation}
	n_{\mathsf{in}} = 2\,{\cal{N}}^2\left[\left( 1 + l^2\right)n_0 + l\kappa\gamma\right]\,, \label{ninA}
\end{equation}
and $n_0 = \alpha^2 + \sinh^2 r$ is the average photon number in the component state $|\alpha,\xi\rangle$, with
\begin{equation}
	\gamma = 2\kappa\left\{\sinh^{2}(r)-\alpha^{2}\left[\sinh(4r)\cos\theta+\cosh(4r)\right]\right\}.\label{expr_gamma}
\end{equation}

The most challenging terms to compute are of the kind $\langle-\alpha,\xi|U|\alpha,\xi\rangle$
(the $A$ subscripts have been omitted) and can be obtained considering that
\begin{equation}
	\langle-\alpha,\xi|U|\alpha,\xi\rangle  =  \langle 0| U'\,D\left(\zeta\right)| 0\rangle.\label{auxiliareq}
\end{equation}
Here  $U' = S^\dagger(\xi) D^\dagger(-\alpha) U D(-\alpha) S(\xi)$, 
$D(\zeta) = \exp\left(\zeta\hat{a}^\dagger - \zeta^* \hat{a}\right)$ 
and $\zeta = 2\alpha\cosh r+2\alpha e^{i\theta}\sinh r$.
In this way, the QFI can be computed straightforwardly term by term. After some
manipulations we obtain:
\begin{equation}
	H\left(\phi\right)=4\,{\cal{N}}^2\left[\left(1+l^{2}\right)\gamma_{1}+l\kappa\gamma_{2}\right]-
	\left(n_{\mathsf{in}}\right)^{2} \,,
\end{equation}
where we have defined
\begin{eqnarray}
	\gamma_{1} & \equiv & _{A}\langle{\alpha,\xi}|G_{A}^{2}|{\alpha,\xi}\rangle_{A}={}_{A}\langle{-\alpha,\xi}|G_{A}^{2}|{-\alpha,\xi}\rangle_{A}\\ 
	\nonumber
	& = & \alpha^{4}-\alpha^{2}\left[1+\sinh(2r)\cos\theta\right]+\frac{1}{2}\left(4\alpha^{2}-1\right)\cosh(2r)+\frac{3}{8}\cosh(4r)+\frac{1}{8}\,,
\end{eqnarray}
and
\begin{eqnarray}
	\gamma_{2} & \equiv & 2\mathrm{Re}\left[_{A}\langle-\alpha,\xi|G_{A}^{2}|\alpha,\xi\rangle_{A}\right]\\ \nonumber
	& = & \frac{1}{4}\kappa\left\{ 2\alpha^{2}\left[\alpha^{2}\left(1+2\sinh^{2}(4r)\cos(2\theta)+3\cosh(8r)\right)-2(\sinh(2r)+3\sinh(6r))\cos\theta\right.\right.+\\ \nonumber
	&  & \left.-6\cosh(6r)\right]+4\cosh(2r)\left[4\alpha^{2}\left(2\alpha^{2}\cosh(4r)+1\right)\sinh(2r)\cos\theta-\alpha^{2}-1\right]+\\ \nonumber
	&  & \left.+\left(8\alpha^{2}+3\right)\cosh(4r)+1\right\} \,.
\end{eqnarray}
Needless to say, if we let $l=0$ we re-obtain the following equation derived in reference \cite{monras06}
\begin{equation}
	H\left(\phi\right)_{l=0}=4\alpha^2 \left[\cosh(2r)-\sinh(2r) \cos\theta\right] + \cosh(4r)-1\,.\label{qfilzero}
\end{equation}
We now re-parametrize the QFI as a function of $n_{0}$ and the ``squeezing fraction of the component state 
$|\alpha,\xi\rangle$", which is $\beta = \sinh^2 r/n_0$. In this approach, we should interpret the parameters $n_{0}$
and $\beta$ as just two auxiliary parameters that will be useful to compare the results of this work with 
the non-entangled case \cite{douglas14}.

The energy of the input state depends on the parameters $n_{0}$ and $\beta$ of the component state 
$|\alpha,\xi\rangle$ as well as on the interpolating parameter $l$.
For this reason, we represent the QFI as a function of the input average
photon number $n_{\mathsf{in}}$ [Eq.~(\ref{ninA})] of the entangled state 
$|\Psi_{l}\rangle$ and as a function of the parameter $\beta$. It is not an easy task to
algebraically invert Eq.~(\ref{ninA}) to obtain $n_{0}$ explicitly
as function of $n_{\mathsf{in}}$, so we do this numerically, adjusting
the value of $n_{0}$ in order to get the desired input photon number
$n_{\mathsf{in}}$.
In Fig.~\ref{fig:QFI-ZERO-Eta} ({\it top}) we plot the QFI as a function of the
squeezing fraction $\beta$. The optimal probe state is
the one capable of reconciling the gains due to entanglement without
losing the gains due to squeezing. We notice that when $l>0$ we
have a ``squeezing fraction'' $\beta < 1$ for which the QFI is greater
than the one obtained with non-entangled states. However, when $l<0$, although
we have a state that may be a maximally entangled quasi-Bell state 
when $l\rightarrow-1$, we notice that we do not have any increase for the QFI. 
Thus we conclude that the best strategy is to spend all the energy in squeezing the state. 
To understand this phenomenon, we analyze the parameter $n_{0}$ that the component
state $|\alpha,\xi\rangle$ must have in order to let the input photon
number be the available value $n_{\mathsf{in}}$ [Fig.~\ref{fig:QFI-ZERO-Eta} ({\it middle})]. 

Because the energy $n_{\mathsf{in}}$ increases
for $l\rightarrow-1$, we must reduce $n_{0}$ to keep the energy
$n_{\mathsf{in}}$ fixed while we change $l$. This implies a
reduction of the QFI when $l\rightarrow-1$, unless we make $\beta=1$
and the optimal state is not entangled. In the graphs for the QFI and $n_{0}$, we have used the optimal 
squeezing angles $\theta$ ($\theta_{\mathsf{opt}}$), which are plotted in Fig.~\ref{fig:QFI-ZERO-Eta} ({\it bottom})
as a function of $\beta$.

\begin{center}
	\begin{figure*}[h]
		\begin{centering}
			\includegraphics[height=5.3cm]{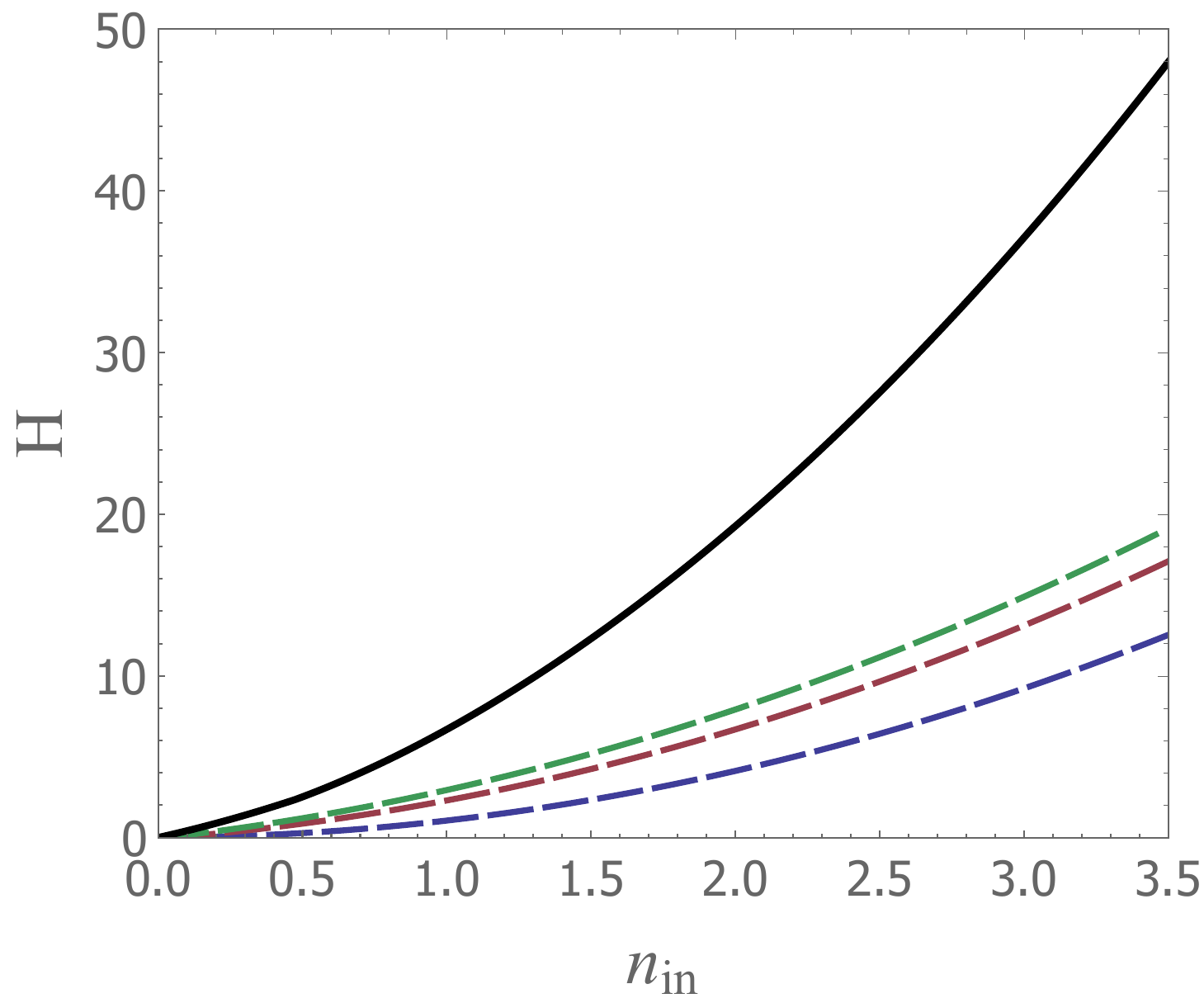}
			\par\end{centering}
		
		\caption{\label{fig:QFI_Comparison}QFI for small phase estimation as a function of the input energy,
			for some commonly used states. From bottom to top: N$00$N state, Caves' state, optimal Gaussian state, 
			and the optimal quasi-Bell state.}
		
	\end{figure*}
\end{center}

The phase estimation performance of the squeezed quasi-Bell state $|\Psi_{l}\rangle$ may be compared to the performance of 
other well-known states in a straightforward way, for instance, by juxtaposing the curves of QFI relative to
different probe states having the same available input energies. This is shown in Fig. \ref{fig:QFI_Comparison}, 
where we have plotted the QFI as a function of the input energy, $n_{\mathsf{in}}$, for some known states. 
For instance, an early proposition by Caves \cite{caves81} involves the mixing of a coherent state $|\alpha\rangle_{1}$ 
and the squeezed vacuum state $|\xi\rangle_{2}$ in a $50:50$ beam splitter at the preparation stage, 
which would result in the following probe state (Caves' state)
\begin{equation}
  |\psi_C\rangle = \exp\left[-i\frac{\pi}{4}\left(\hat{a}_1^\dagger\hat{a}_2 + \hat{a}_2^\dagger\hat{a}_1\right)\right]
  |\alpha\rangle_{1} |\xi\rangle_{2}.\label{cavesstate}
\end{equation}
It is clearly seen in Fig. \ref{fig:QFI_Comparison} that the optimal quasi-Bell state considerably outperforms the N$00$N state, 
Caves' state in Eq. (\ref{cavesstate}) as well as the optimal Gaussian state (squeezed vacuum state $|\xi\rangle = S(\xi) |0\rangle$).
We recall that the optimal Caves' state is such that half of the input intensity is provided by the coherent state and 
the other half by the squeezed state.

\subsection{Phase estimation with a linear unitary disturbance}

We now consider the estimation of phase under a linear unitary disturbance $Q_A = \hat{a}_A + \hat{a}_A^\dagger$. 
The corresponding evolution operator is \cite{douglas14,pasquale13}
\begin{equation}
	U_{\phi,\eta}=\exp\{-i(\phi G_{A}+\eta Q_{A})\},
\end{equation}
where $G_{A}=\hat{a}_{A}^{\dagger}\hat{a}_{A}$ and $\eta$ is the strength of the disturbance. Again, we will be using the 
partially entangled states $|\Psi_{l}\rangle$. As we are dealing with pure states, we may follow the general procedure 
introduced in \cite{pasquale13} to compute the QFI using the expression
\begin{equation}
H\left(\phi\right)=4\left[\langle\Psi_{l}|\bar{G}_{A}^{2}|\Psi_{l}\rangle-\left(\langle\Psi_{l}|\bar{G}_{A}|\Psi_{l}\rangle\right)^{2}\right]\,.
\end{equation}
Here, the average generator of the evolution of mode $A$ is given by
\begin{equation}
\bar{G}_{A}=G_{A}+\frac{\eta}{\phi^{2}}\left(\phi Q_{A}+2\eta\right)\left(1-\frac{\sin\phi}{\phi}\right)+\frac{\eta}{\phi^{2}}P_{A}\left(\cos\phi-1\right)\,,
\end{equation}
with $P_{A}=-i\left(\hat{a}_{A}-\hat{a}_{A}^{\dagger}\right)$. Now using Eq.~(\ref{auxiliareq}) and doing some algebraic manipulations for 
the other terms, we obtain an analytical expression for the QFI. Due to its large size, though, we opted to omit that expression here.

\subsubsection{Optimal state for the estimation of a very small phase $\phi$}

\begin{figure}[h]
	\begin{centering}
		\begin{tabular}{ccc}
			\includegraphics[height=3.8cm]{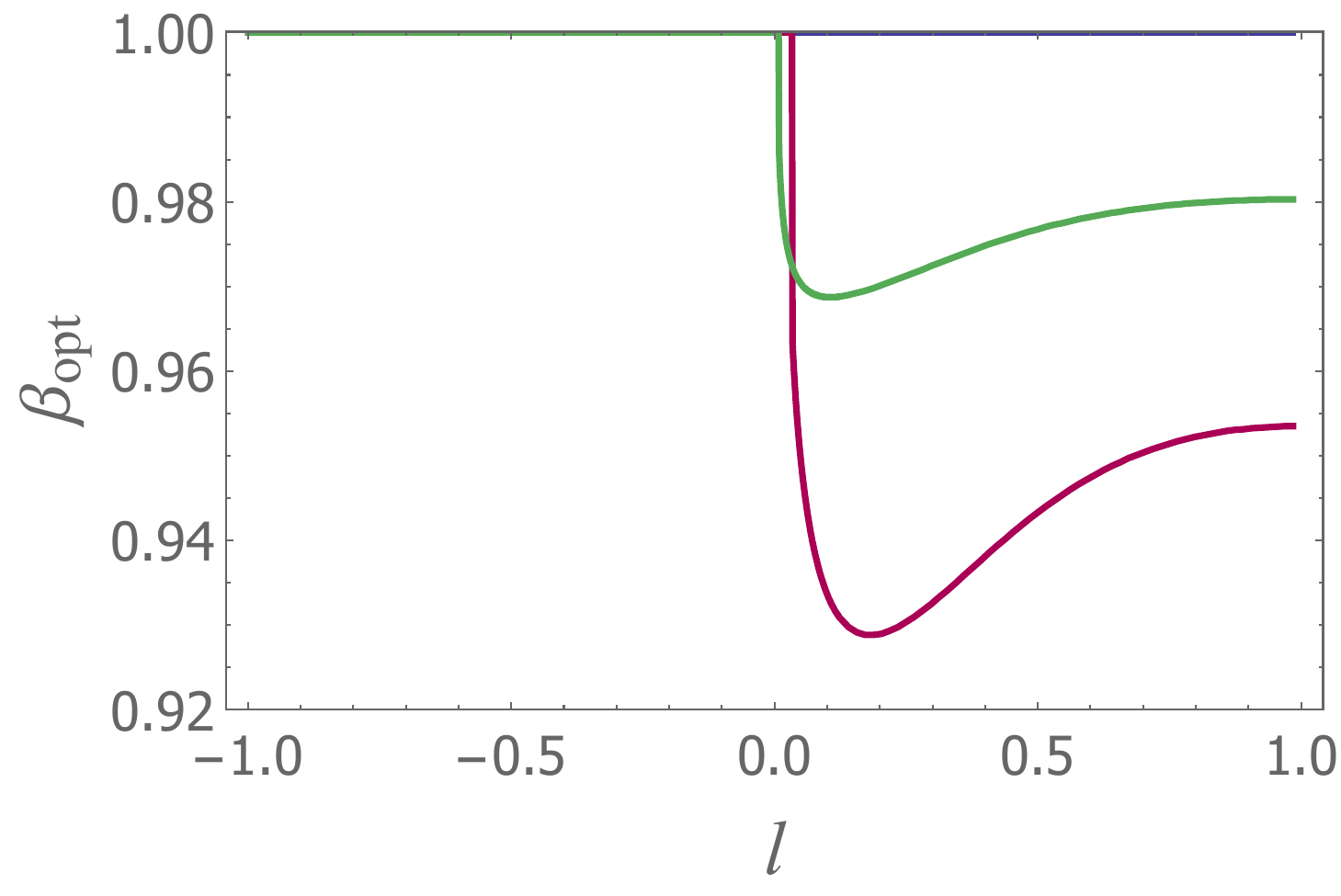} &  & 
			\includegraphics[height=3.8cm]{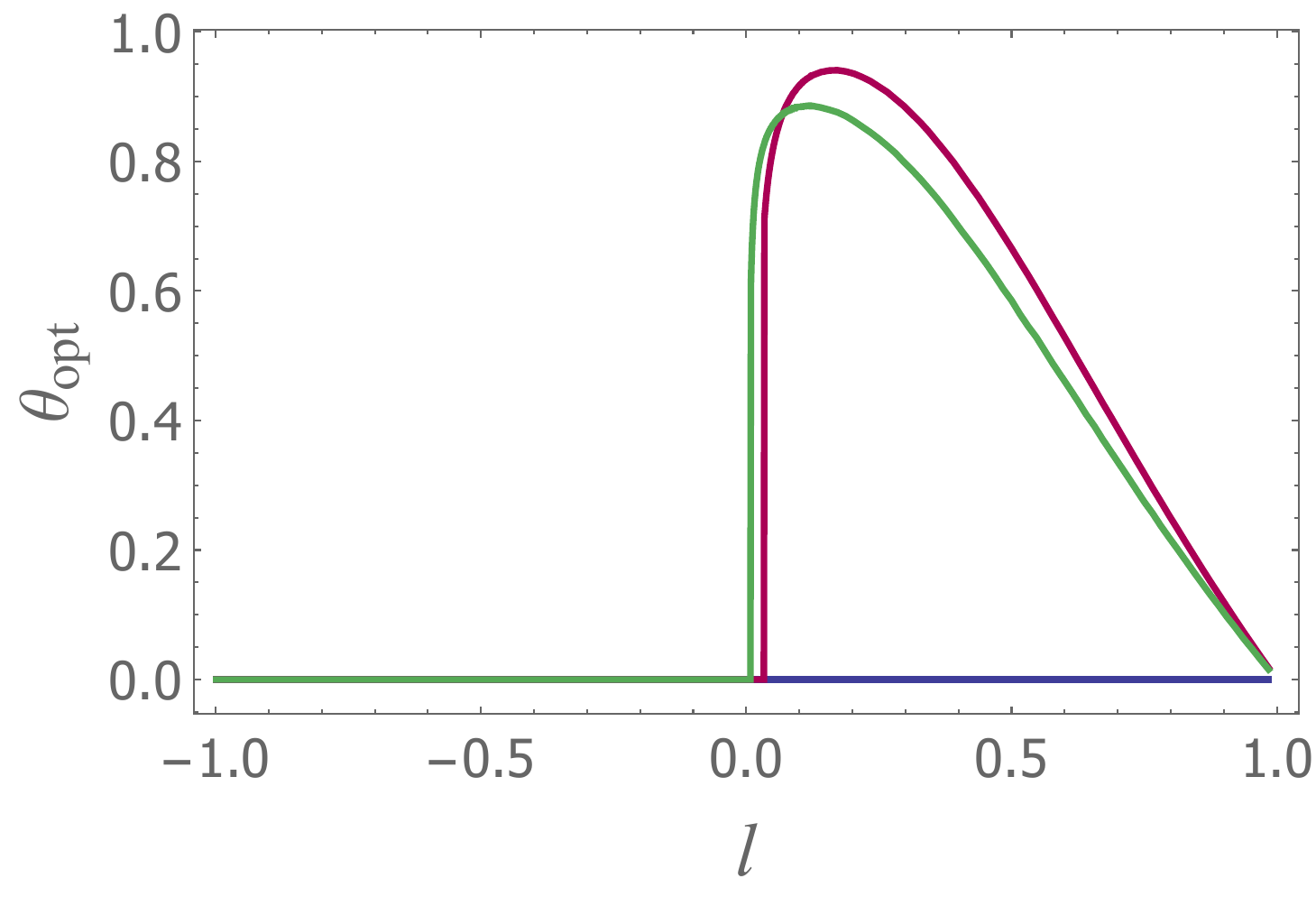}\tabularnewline
		\end{tabular}
		\par
		\begin{tabular}{ccc}
			& \includegraphics[height=5cm]{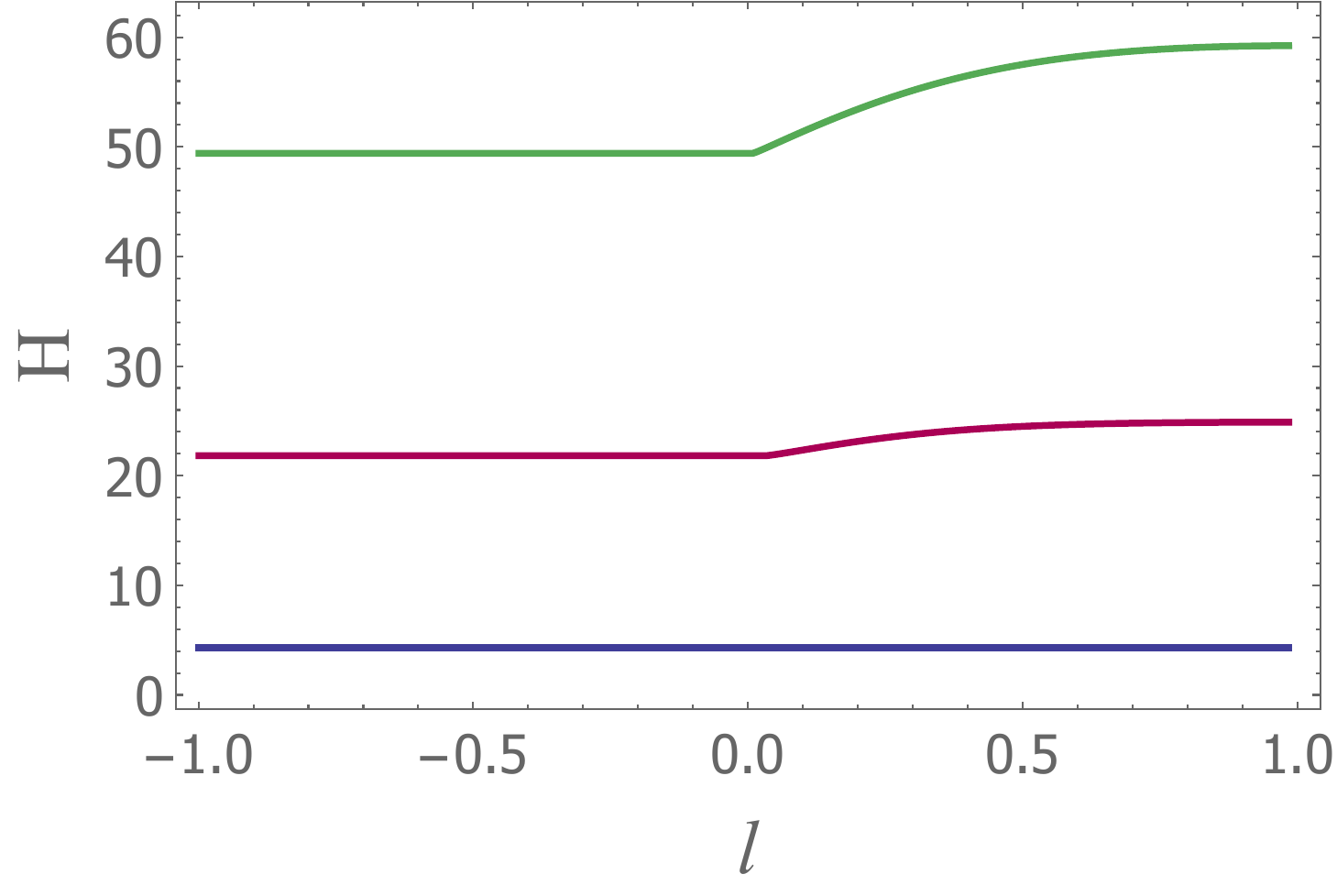} & \tabularnewline
		\end{tabular}
		\par
	\end{centering}
	
	\caption{\label{fig:QFI-functionOf-L}Optimal parameters
		``squeezing fraction'' ${(top-left)}$ and ``squeezing angle'' ${(top-right)}$
		of the component state as a function of $l$. ${(bottom)}$ QFI $H$ as a function of the interpolation parameter
		$l$ for the optimal input states. We have $n_{\mathsf{in}}=\{0.4,\,2.0,\,3.6\}$
		from bottom to top in the $H$ plot. The case $n_{\mathsf{in}}=0.4$ is the almost invisible blue line at 
		$\beta_{\mathsf{opt}}=1.0$ and at $\theta_{\mathsf{opt}}=0.0$. For a higher $n_{\mathsf{in}}$ we have $\beta_{\mathsf{opt}}\neq1.0$ and 
		$\theta_{\mathsf{opt}}\neq0.0$ for $l>0$. (\textbf{the colors identify the curves in the $\beta_{\mathsf{opt}}$ and $\theta_{\mathsf{opt}}$ 
			plots according to the $H$ plot}). For all the plots we have $\eta=1$.}
\end{figure}

After computing the limit $\phi\rightarrow0$, we want to find the quasi-Bell state which allows
the maximum precision for the phase estimation. In order to do so, we
keep the input energy $n_{\mathsf{in}}$ fixed. We consider different
values for the interpolation parameter $l$ and then we maximize the
QFI as being a function of the parameters $\beta$ and $\theta$. In Fig.~\ref{fig:QFI-functionOf-L}
we plot the optimal ``squeezing fraction'' $\beta_{\mathsf{opt}}$,
the optimal squeezing angle $\theta_{\mathsf{opt}}$, as well as the maximized QFI $H$, as a function
of the parameter $l$. Each curve corresponds to a fixed input average photon number $n_{\mathsf{in}}$.

We have verified that if the parameter $l$ is negative, the optimal input
state is not an entangled state, and the mode $A$ of this state corresponds
to what has been previously found \cite{douglas14}, i.e.,
the QFI is given by 
\begin{equation}
	H = 8\left(n_0 + n_0^2\right) + \left(2n_0 + 2\sqrt{\left(n_0 + n_0^2\right)} + 1\right)\eta^{2}.
\end{equation}

\begin{center}
	\begin{figure*}[h]
		\begin{centering}
			\includegraphics[height=5.3cm]{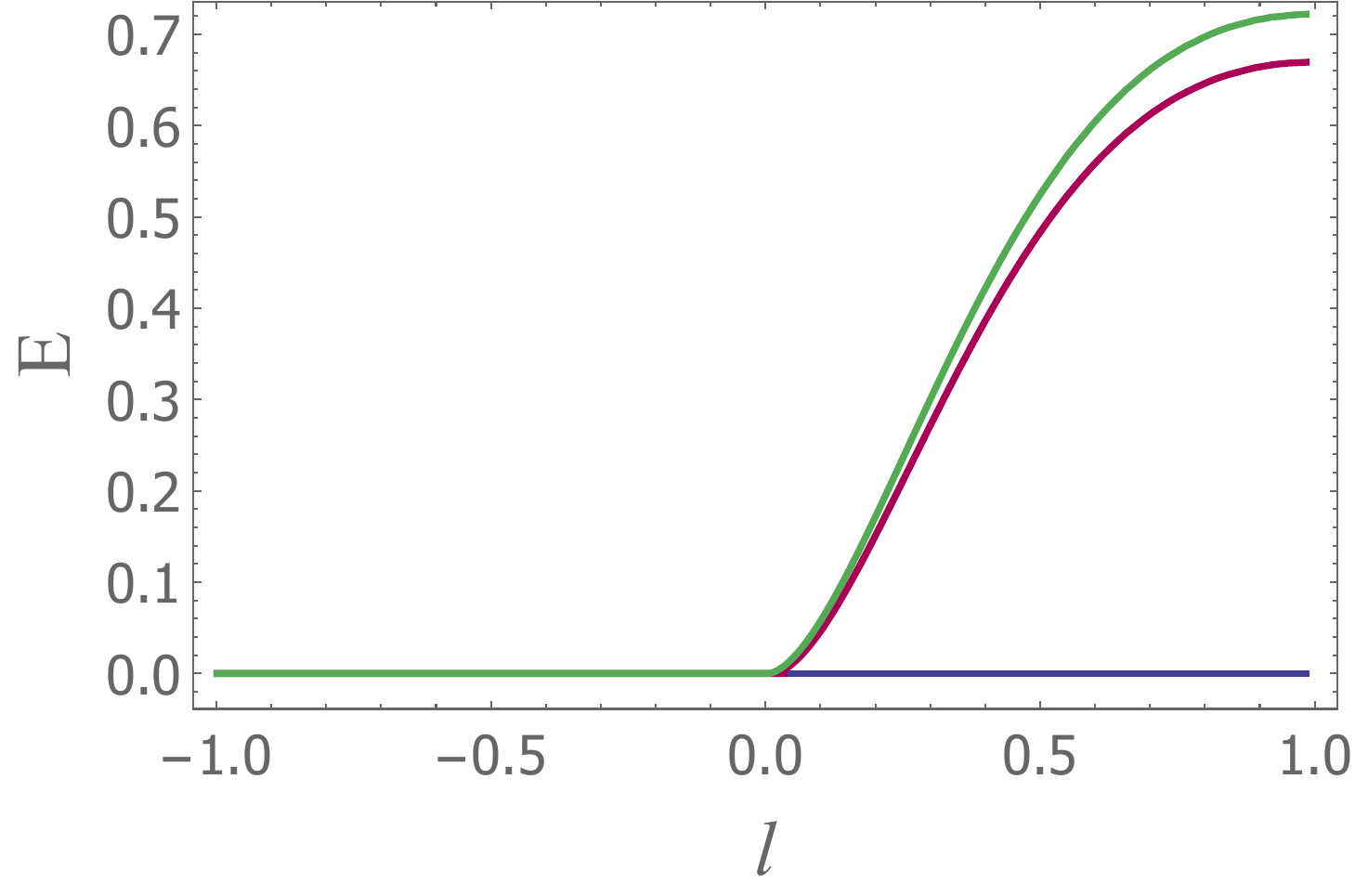}
			\par\end{centering}
		
		\caption{\label{fig:Emaranhamento-Sonda-otima}Entanglement of the optimal
			probe state as a function of the interpolation parameter $l$. As in
			the previous plots, $n_{\mathsf{in}}=\{0.4,\,2.0,\,3.6\}$
			from bottom to top. The first curve, $n_{\mathsf{in}}=0.4$, is hidden at $E=0.0$.}
		
	\end{figure*}
\end{center}

We remind that in the case of phase estimation with single-mode states we have
$\beta_{\mathsf{opt}}=1$. If $l$ increases from $0$ to $1$,
though, the QFI increases if there is enough energy. In this range
of values for the energy, increasing $\beta$ past $\approx0.8$ leads
to a reduction of entanglement. This is because the components of the quasi-Bell
state become two squeezed vacuum states, and the overlap $\kappa=\langle{\alpha,\xi}|{-\alpha,\xi}\rangle$ 
increases \cite{hirota}. For this reason, $\beta_{\mathsf{opt}}$ is not equal to $1$, which 
reconciles the gains due to entanglement with the
gains due to squeezing for the phase estimation. In Fig.~\ref{fig:Emaranhamento-Sonda-otima}
we plot the entanglement of the optimal probe state as a function of the interpolation
parameter $l$. We draw attention to the fact that even when we take
$l\neq0$, entanglement is not imposed, because the state is not entangled if
$\beta_{\mathsf{opt}}=1$, as it happens for $l<0$. Moreover, in the case of a single-mode
squeezed state (when there was no additional parameter $l$)
$\beta_{\mathsf{opt}}=1$ was indeed the optimal value for the ``squeezing fraction". 
This means that the phase estimation with a linear unitary disturbance could be upgraded
by the use of entangled states.

\begin{figure*}[h]
	\begin{centering}
		\includegraphics[height=5.3cm]{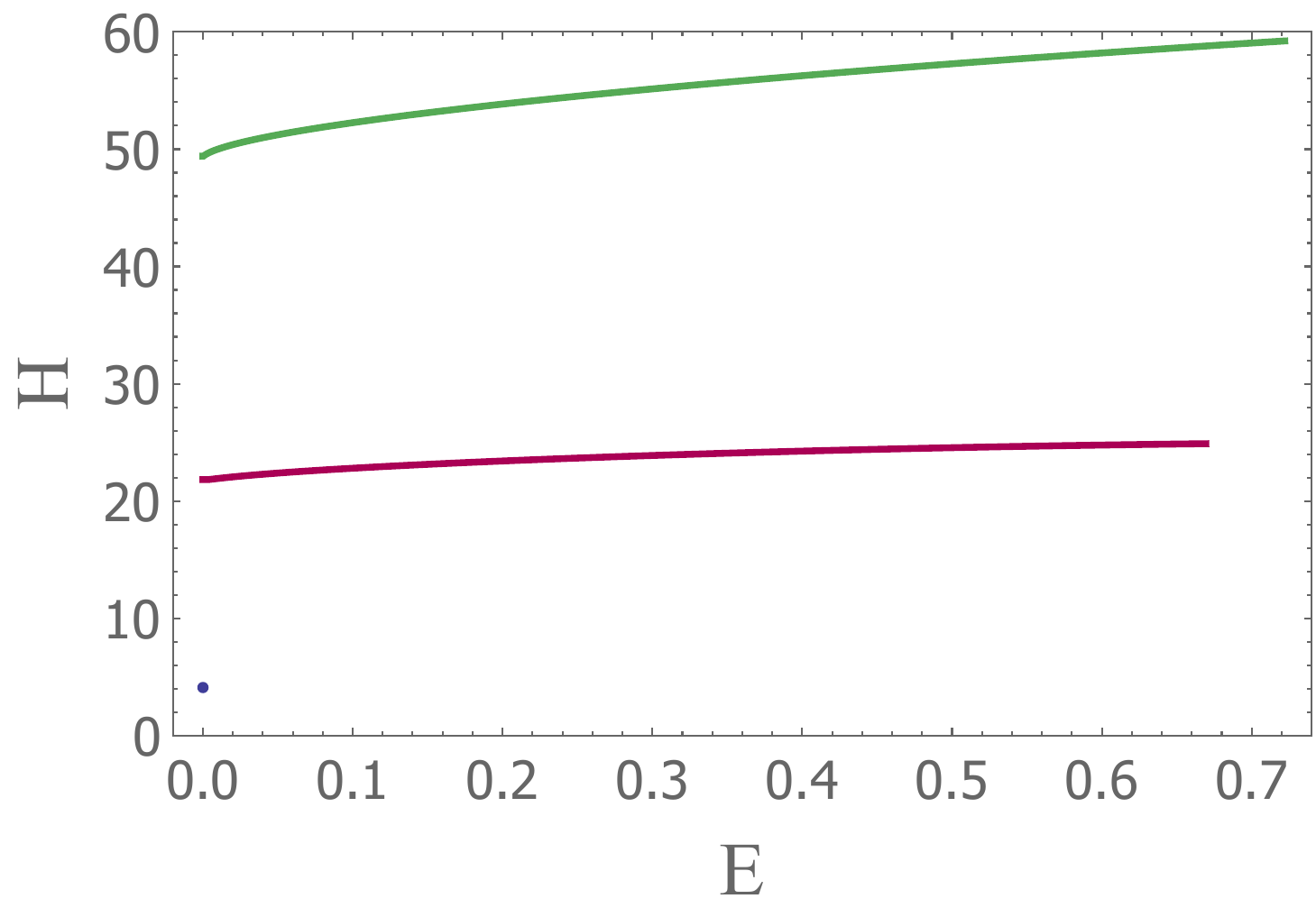}
		\par\end{centering}
	
	\caption{\label{fig:QFI_functionOf_E}QFI $H$ obtained from probe states with entanglement $E$ 
		for $l$ varying from $0$ to $1$ (the lines are drawn from left to right as $l$ increases). All the other parameters 
		were optimized in order to maximize the QFI for each value of $l$, maintaining the input photon number $n_{\mathsf{in}}$ 
		fixed. The maximal reached entanglement (at $l=1$) is different for each state and do not reach 1, that is why the plots 
		end at different values of $E$. We have $n_{\mathsf{in}}=\left\{0.4,\,2.0,\,3.6\right\} $ from bottom to top. 
		For $n_{\mathsf{in}}=0.4$ the plot is just a blue dot at $E=0$, $H=4.0$. We have $\eta=1$ for all the plots.}
	
\end{figure*}

The QFI is gradually increased when we allow the state
to be more and more entangled (increasing $l$), so it seems natural
to look for a more direct relation between the entanglement of the
quasi-Bell states and the resulting QFI. When we found the optimal parameters 
$\beta$ and $\theta$ for the
input state by maximizing the QFI, we removed the dependence of the
QFI on those parameters. We are now able to observe the dependence of the
QFI on the interpolation parameter $l$, which fixes the amount of
entanglement of the input state. Because both the QFI and the entanglement
are monotonic functions of $l$, for $l>0$, we can
represent the QFI directly as a function of the entanglement for this
(positive) $l$ semi-axis. In Fig.~\ref{fig:QFI_functionOf_E} we notice
that the QFI increases monotonically as the entanglement of the probe
state increases, showing that entanglement is a resource for quantum
phase estimation even if there is a unitary disturbance in the system.

\subsubsection{The effect of disturbance on the phase estimation with entangled states}

The QFI is an increasing function of the parameter $\eta$ of the
disturbance because there is an energy increase parameterized by $\eta$ during the transformation \cite{douglas14}.
We can see this by computing the average photon number in the output state, with the help of Eq. (\ref{auxiliareq}):

\begin{equation}
	n_{\mathsf{out}}=\left\langle \Psi_{l}\right|U_{\phi,\eta}^{\dagger}
	\,n\,U_{\phi,\eta}\left|\Psi_{l}\right\rangle=n_{\mathsf{in}}+\eta^2 \,.
\end{equation}

\begin{figure*}[h]
	\begin{centering}
		\includegraphics[height=6cm]{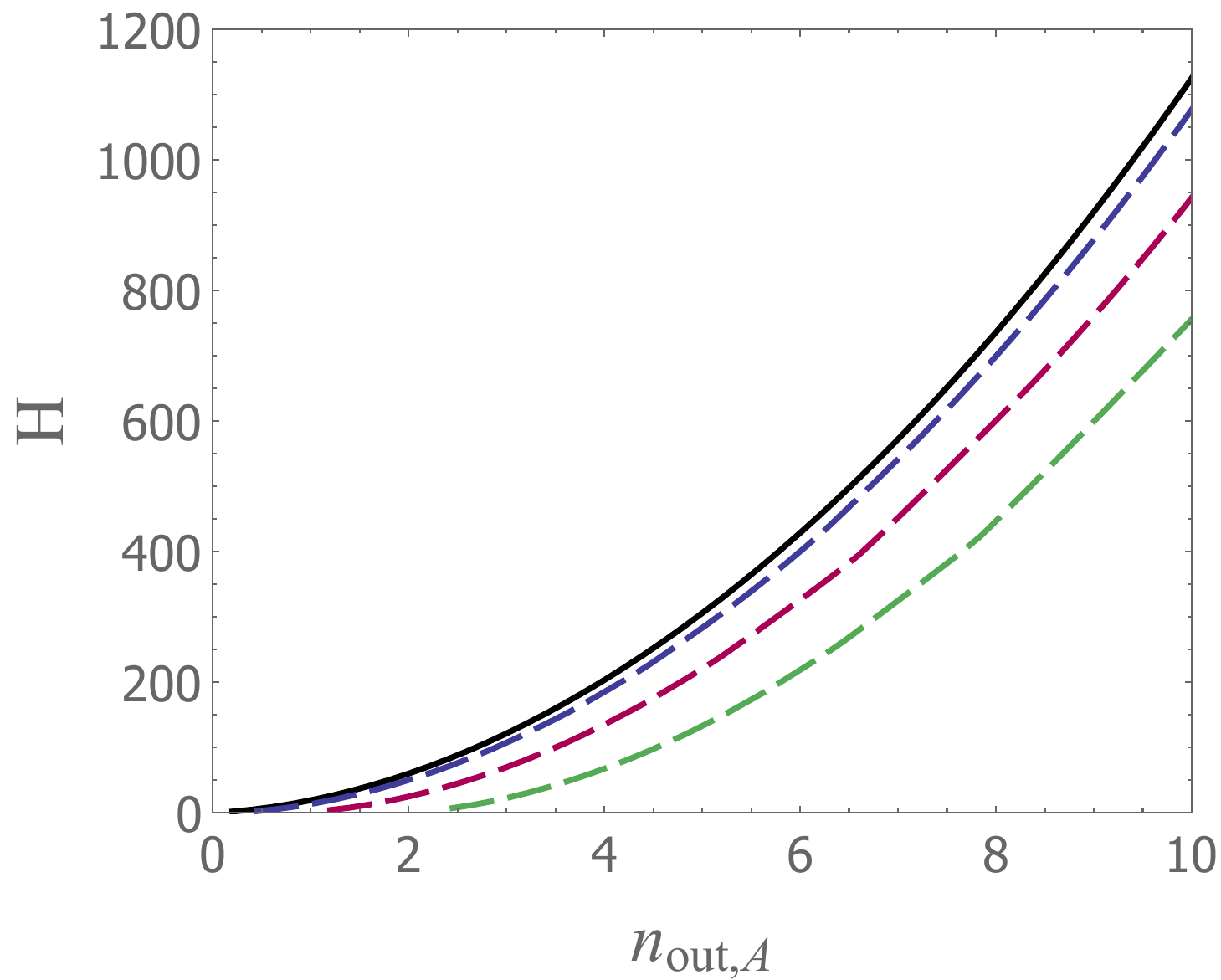}
		\par\end{centering}
	
	\caption{\label{fig:QFI-functionOf-nOut-Entanglement}QFI $H$ as a function
		of the average photon number $n_{\mathsf{out}}$ on the mode $A$
		of the output state of the transformation for $l=1$. The dashed lines are, 
		from top to bottom, for $\eta=\left\{ 0.5,\,1.0,\,1.5\right\}$. 
		For comparison, the solid black line corresponds to $\eta = 0$ \cite{douglas14}.}
	
\end{figure*}

In order to perform a more precise analysis of how the QFI depends on the disturbance,
we plot in Fig.~\ref{fig:QFI-functionOf-nOut-Entanglement} the QFI
as a function of the average photon number in mode $A$ for $l=1$. 
Of course for $l=0$ we re-obtain the previous results for single-mode Gaussian
states \cite{douglas14}.

We note that the presence of the disturbance $\eta$ affects the phase estimation
also when the probe state is entangled, if the total available energy (input state + transformation)
is taken into account. However, we observe that the QFI may attain larger values than in the non-entangled 
case \cite{douglas14}, showing once more the advantages of using entangled states for quantum phase estimation
even in the presence of a unitary disturbance.

Finally, we may analyze the feasibility of the adjustment of the parameters
that were optimized along this work. Firstly we remark that the plot
in Fig.~\ref{fig:QFI-functionOf-nOut-Entanglement} corresponds
to the behavior of the left ends ($l=1$) of the plots in Fig.~\ref{fig:QFI-functionOf-L} ({\it bottom}),
when we consider higher energies. For $l=1$, the value of
$\beta_{\mathsf{opt}}$ is well defined, but it may be hard to reach it
for a total input photon number larger than $n_{\mathsf{in}}\sim 4$. This is
because, in this case, we have $0,98\lesssim\beta_{\mathsf{opt}}<1$
and the sensitivity of the QFI upon a very small variation of $\beta$
around $\beta_{\mathsf{opt}}$ is very high. The plot in Fig.~\ref{fig:QFI-functionOf-nOut-Entanglement}
represents then only a theoretical indication of how entanglement
may be useful for phase estimation.

\subsection{Effect of dissipation}

Quantum systems are generally subjected to losses and decoherence, and this will surely affect the performance 
of any phase estimation protocol. Now we would like to discuss the influence of losses if entangled squeezed states
are used as probe states, depending on the values of the parameters involved. In particular, it would be relevant 
to assess the performance of our scheme for different values of the squeezing parameter ($\beta$) as well as the 
amount of entanglement $(l)$.
For this we are going to use the very general result presented in reference \cite{davidovich11}.
In that work, an upper limit $B$ for the QFI valid for any state of the quantized field is derived, being given by
\begin{equation}
	H \leq B = \left[ \frac{4\mu\langle\hat{n}_A\rangle \langle\Delta\hat{n}^2_A\rangle}
	{(1-\mu)\langle\Delta\hat{n}^2_A\rangle + \mu\langle\hat{n}_A\rangle}\right].\label{upperbound}
\end{equation}
Here $\langle\hat{n}_A\rangle$ is the mean photon number in mode $A$, where the phase-shift is applied. 
The parameter $\mu$ is related to the effective damping 
rate in the system; if $\mu = 0$ we have complete absorption of light, and $\mu = 1$ corresponds to the lossless, ideal case. 
We remark that the upper bound in Eq. (\ref{upperbound}) was derived considering losses in mode $A$ only \cite{davidovich11}.
In the case of having the entangled squeezed state as probe state, the upper bound may be derived using previously obtained results.
The mean photon number variance $\langle\Delta\hat{n}^2_A\rangle$ needed for calculating the upper bound $B$ 
in Eq. (\ref{upperbound}) was computed when we obtained the QFI for phase estimation without disturbance 
[see Eq. (\ref{qfimain})], i.e. $\left\langle \Delta\hat{n}_{A}^{2}\right\rangle =\left\langle 
\Psi_{l}\right|G_{A}^{2}\left|\Psi_{l}\right\rangle -\left(\left\langle \Psi_{l}\right|G_{A}\left|\Psi_{l}\right\rangle \right)^{2}$.
The term $\langle\hat{n}_A\rangle$ is the mean photon number in mode A, which is half the mean photon number of the state given by Eq.~(\ref{ninA}).

\begin{figure*}[h]
	\begin{centering}
	\begin{tabular}{ccc}
		\includegraphics[height=5.0cm]{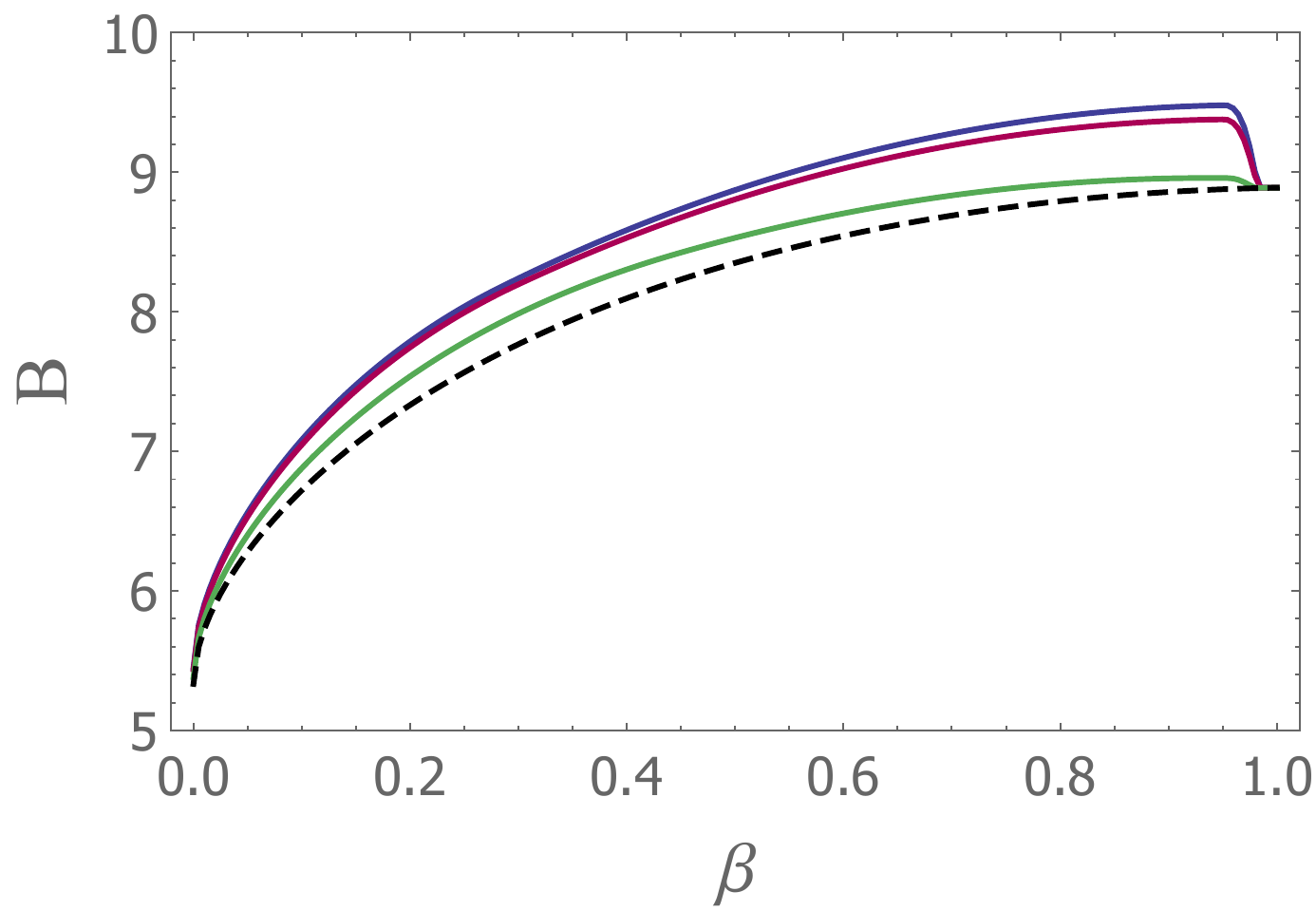} & & \includegraphics[height=5.0cm]{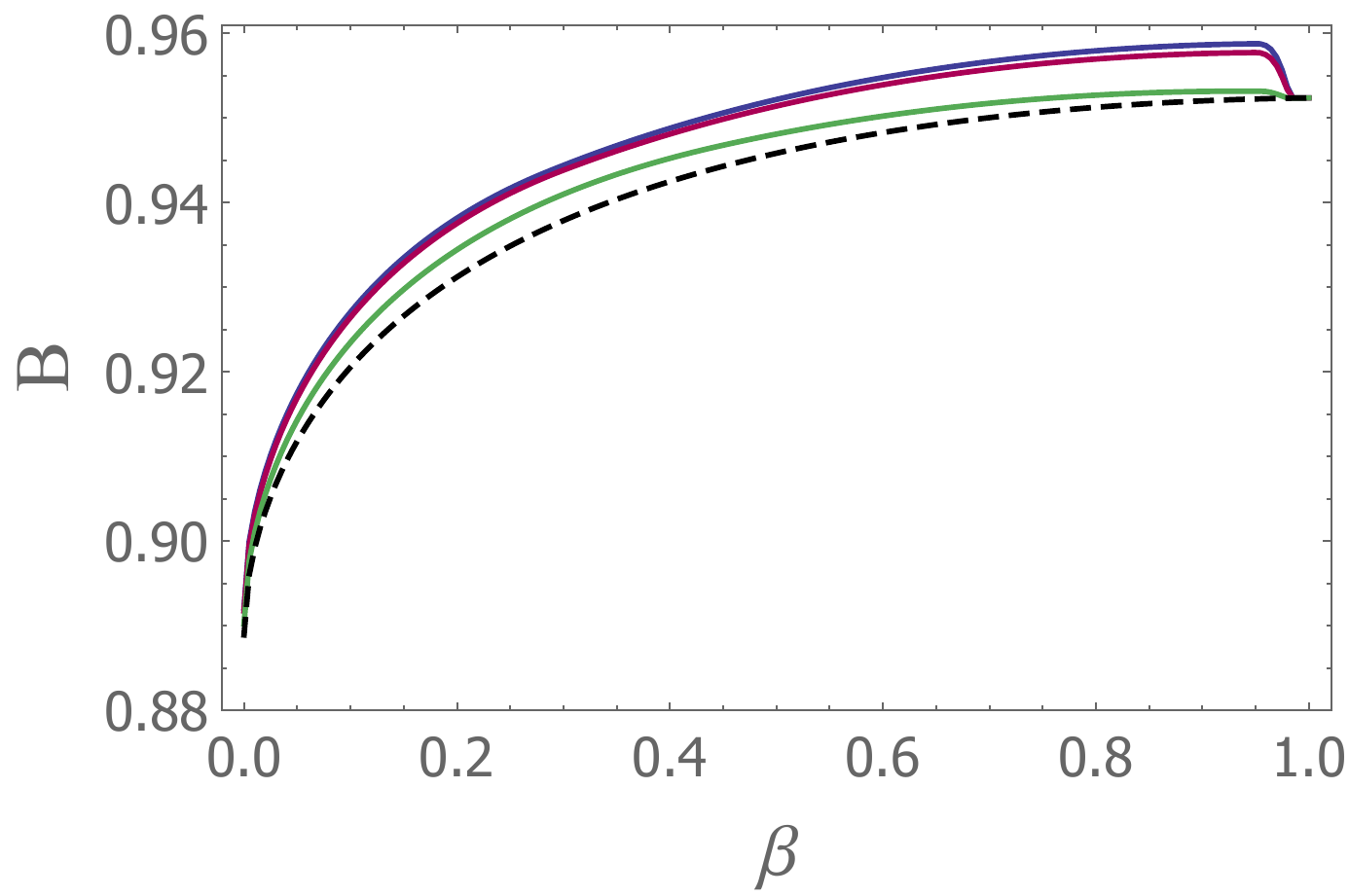}\tabularnewline
		\end{tabular}
			\par
	\caption{\label{fig:boundeta}Upper bound of the QFI for 
		entangled probe states having $l > 0$ and $n_{\mathsf{in}}=1.0$. 
		$(left)$ In the low loss case ($\mu = 0.8$). $(right)$ In the high loss case ($\mu = 0.2$). 
		From top to bottom $l=\left\{ 1.0,\,0.6,\,0.2\right\}$, and the dashed black line 
		corresponds to the non-entangled case ($l=0$).}
		\end{centering}
\end{figure*}

In Fig.~\ref{fig:boundeta} $(left)$ the bound $B$ is plotted as a function of the parameter $\beta$ for $\mu = 0.8$ (low losses). 
We note that the upper bound of the QFI (for a given $\beta$) is basically higher for the entangled 
probe state with $l > 0$ compared to a separable state ($l = 0$). In other words, the maximum possible values for 
the QFI are still larger in the entangled state case, despite the destructive action of dissipation. For higher losses, one 
would expect lower values for the upper bound $B$. This is indeed the case, as shown in Fig.~\ref{fig:boundeta} $(right)$, 
where the bound $B$ is plotted as a function $\beta$ for $\mu = 0.2$ (high losses). We also observe that the curves for different 
values of $l$ become less distinct, which means that as dissipation increases, there is less advantage due to entanglement. 
For instance, the relative difference (with respect to $B$) between the entangled and non-entangled cases may be up to 
$R\approx 7\%$ for $\mu = 0.8$, while for $\mu = 0.2$ it goes down to $R\approx 1\%$. Needless to say that 
according to Eq. (\ref{upperbound}), in the case of complete absorption, $\mu \rightarrow 0$, $B \rightarrow 0$ 
for any values of $l$ and $\beta$. Hence, squeezing is a very useful resource, which allows a better performance
of the ideal phase estimation process and also reduces the detrimental effects of dissipation. In fact a recent work
has experimentally shown that fragile superposition states such ``cat states" may become more robust against losses after
being squeezed \cite{laurat17}.

\section{Conclusion}

In this work we have analyzed the use of the interpolated quasi-Bell states
as input probes for quantum phase estimation. We have found that the use of 
continuous variable entangled states based on squeezed coherent states makes 
possible to increase the precision for phase estimation, specially when the total 
average photon number is not negligible $n_{\mathsf{in}}\gtrsim1$. We have verified 
that the QFI is an increasing function of the interpolation parameter $l$ (for $l>0$), 
which is related to the entanglement contained in the optimal probe state. 

If a unitary perturbation is included in the analysis, we have observed that the larger 
the disturbance parameter $\eta$, the larger will be the energy of the output state, which
increases the QFI. However when we consider all the energy spent in the process 
(including the energy used in the transformation) we found that the
disturbance actually impairs the phase estimation, although entanglement still brings advantages
over the non-entangled case. 
We also highlight that for input states with higher energy, the ``optimal squeezing fraction''
parameter $\beta_{\mathsf{opt}}$ must be finely adjusted in order to maximize
the QFI when $l = 1$. 

We have also discussed the influence of dissipation. We have calculated an upper bound ($B$) for the QFI 
for different values of the interpolation parameter $l$ in a lossy scenario, and found that  
$B_{l>0} \geq B_{l=0}$ (for a fixed value of $\beta$). This means that entanglement still may bring advantages 
in the phase estimation process, even in the presence of dissipation, if Bell-type, entangled squeezed states 
are employed.

\section*{Funding}
Funda\c c\~ao de Amparo \`a Pesquisa do Estado de S\~ao Paulo (FAPESP) grant 2011/00220-5, Brazil.
Conselho Nacional de Desenvolvimento Cient\'\i fico e Tecnol\'ogico (CNPq), INCT - IQ, grants 
2008/57856-6 and 465469/2014-0, Brazil.

\begin{center}
	\begin{figure}[h]
		
		\begin{centering}
			\begin{tabular}{ccc}
				\includegraphics[height=4.6cm]{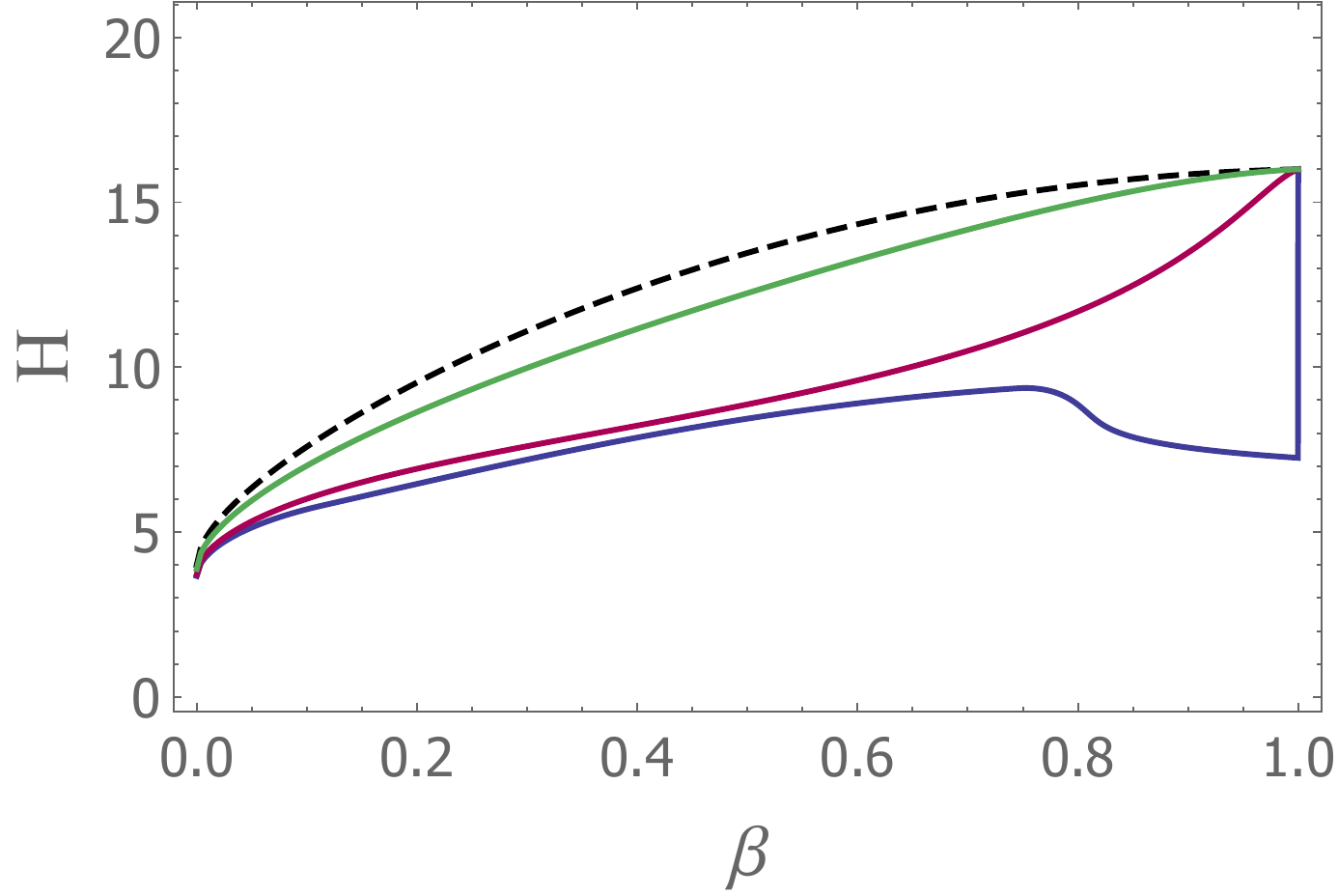} &  & \includegraphics[height=4.6cm]{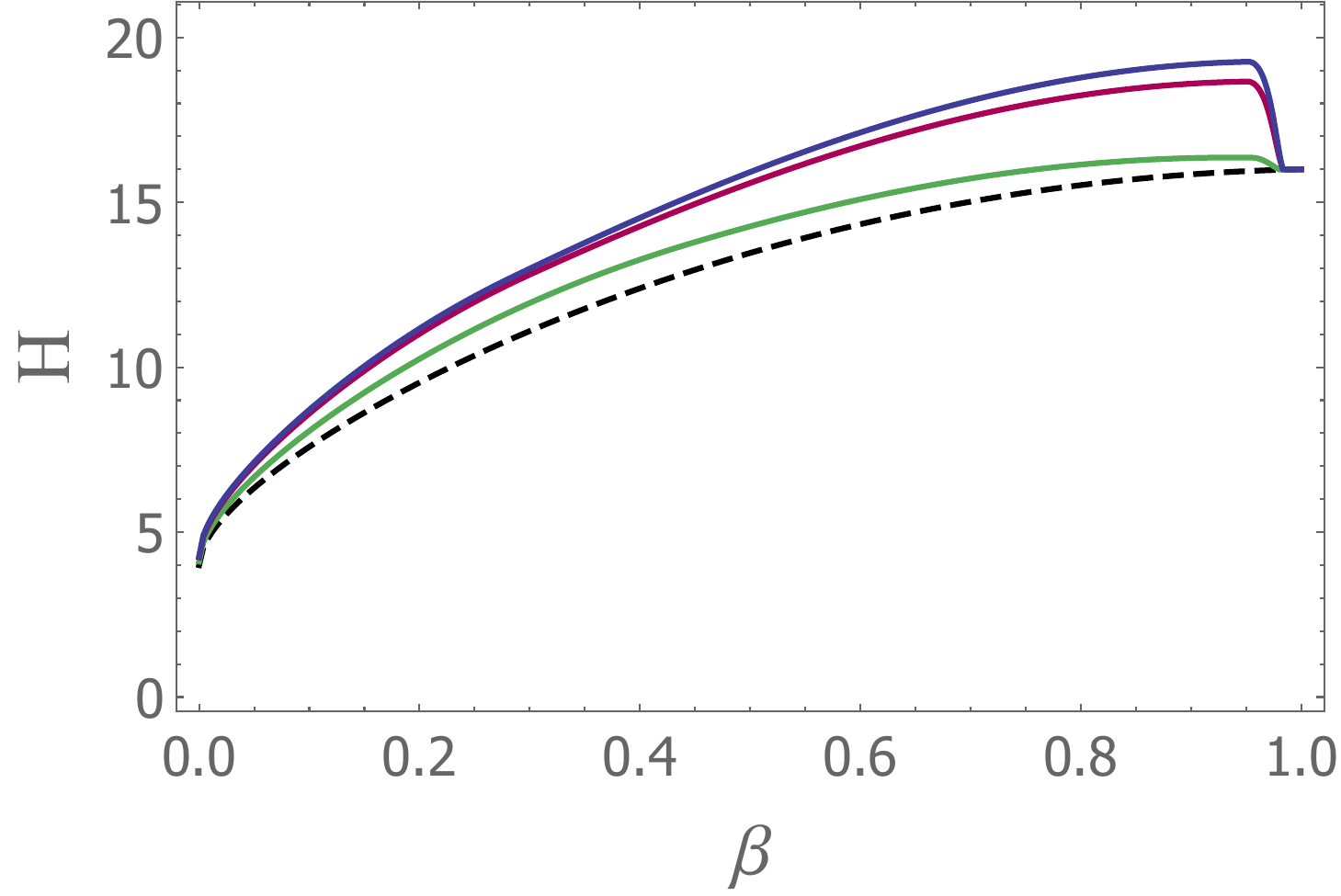}\tabularnewline
			\end{tabular}
			\par
			
			\begin{tabular}{ccc}
				\includegraphics[height=4.6cm]{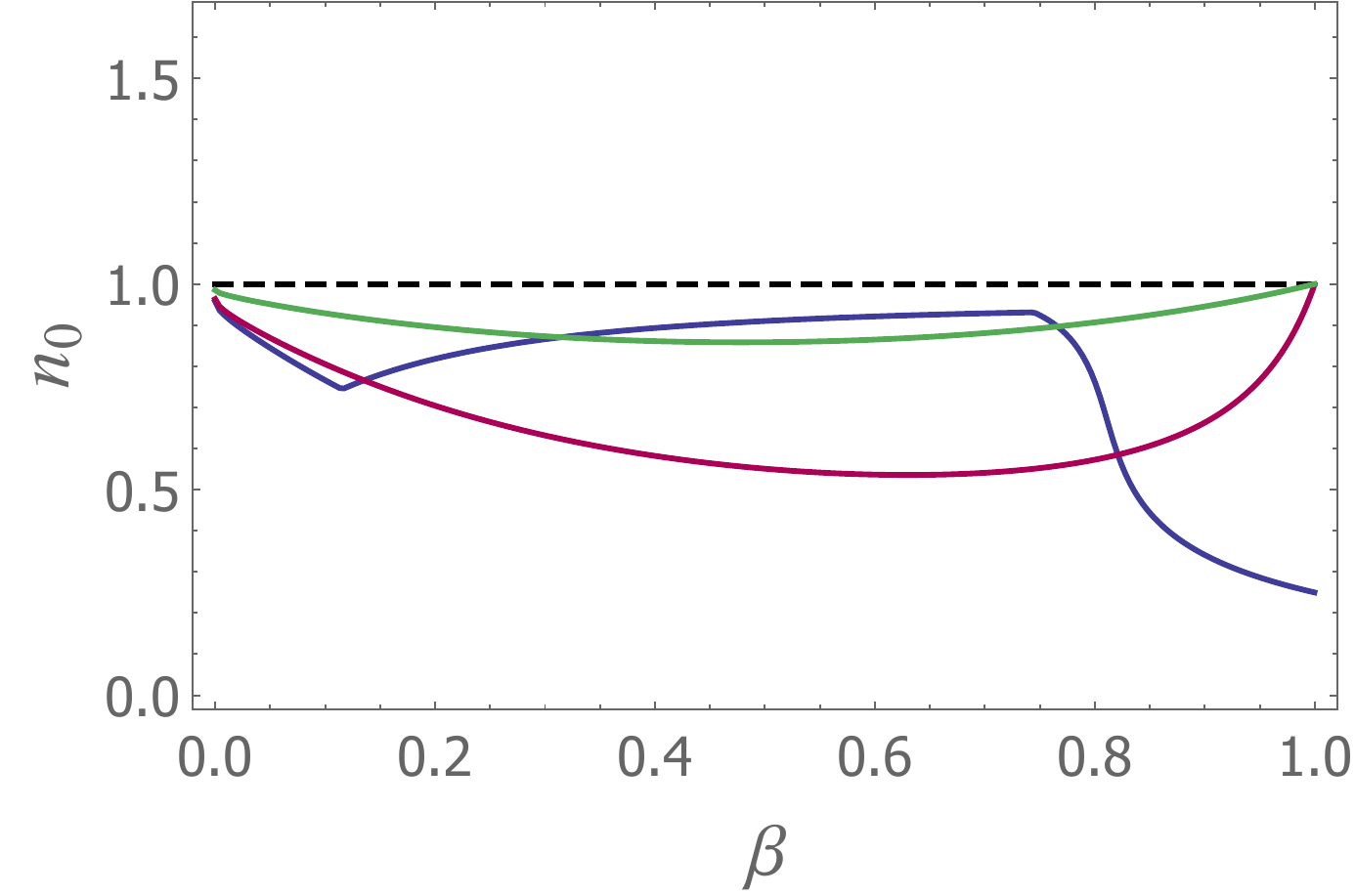} &  & \includegraphics[height=4.6cm]{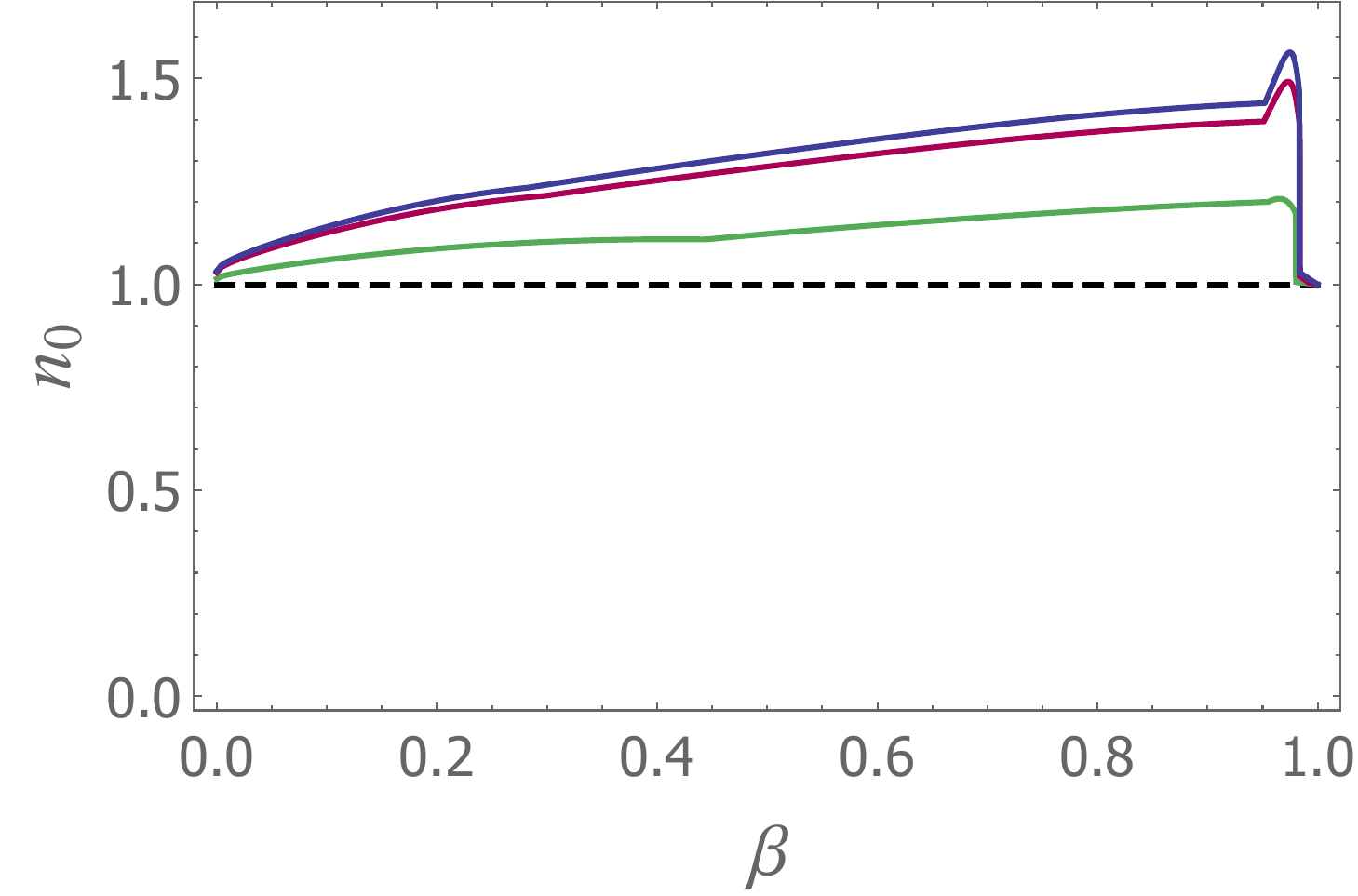}\tabularnewline
			\end{tabular}
			\par
			
			\begin{tabular}{ccc}
				\includegraphics[height=4.6cm]{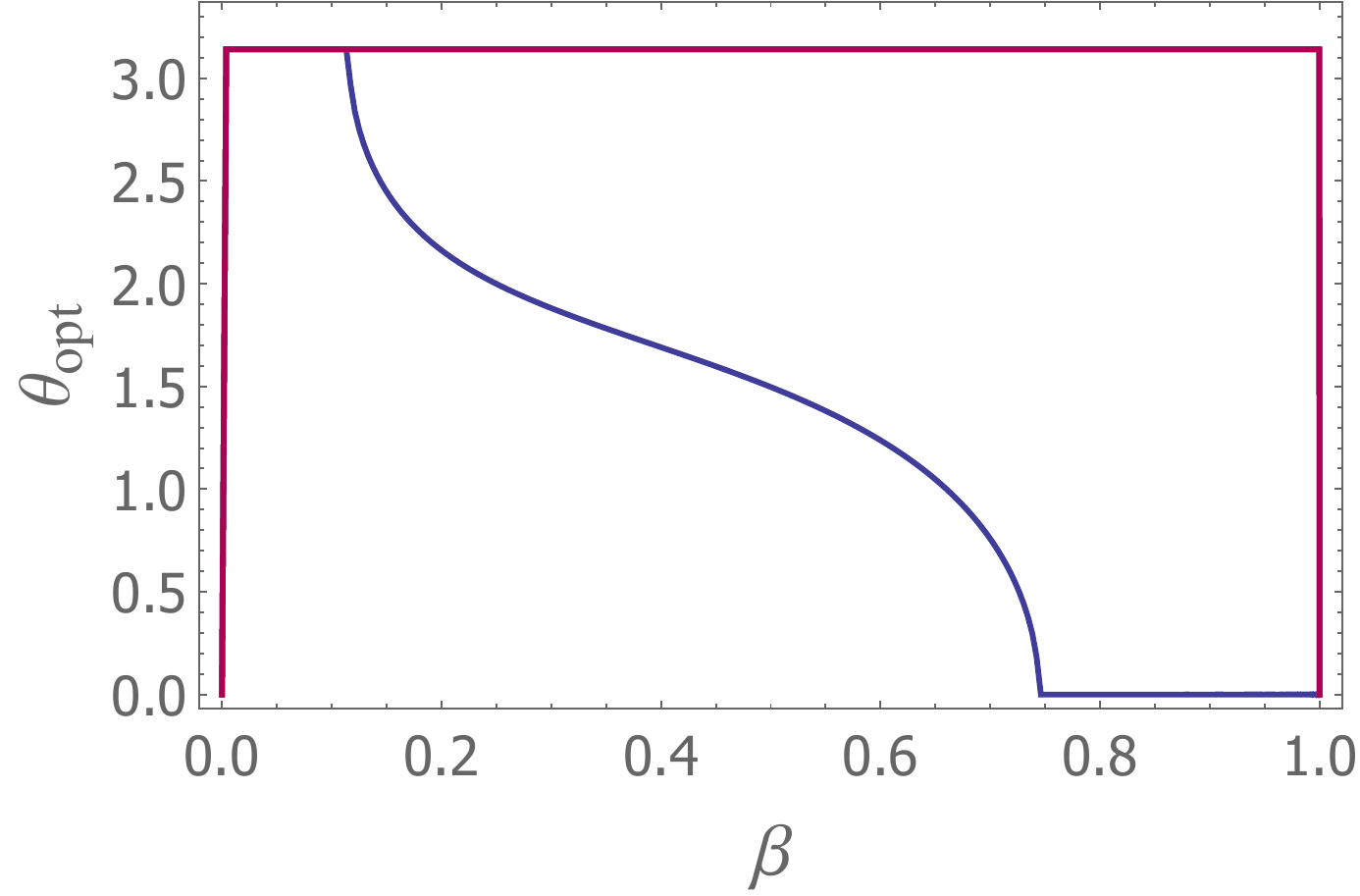} &  & \includegraphics[height=4.6cm]{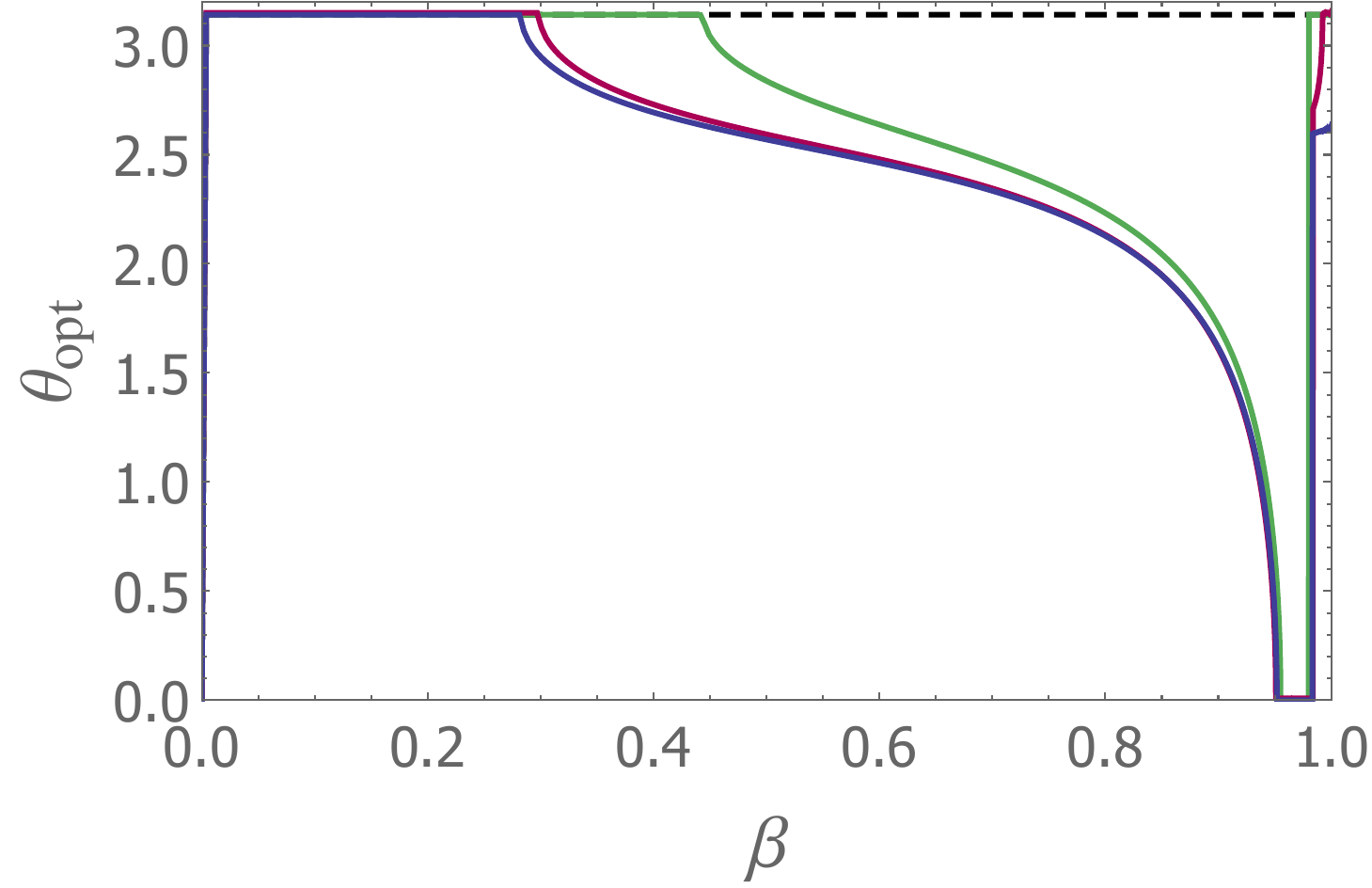}\tabularnewline
			\end{tabular}
			\par
			
		\end{centering}
		
		\caption{\label{fig:QFI-ZERO-Eta}QFI for the phase estimation using entangled
			probes. ${(top-left)}$ Negative $l$ parameter, from bottom to
			top we have $l=\left\{ -1.0,\,-0.6,\,-0.2\right\} $.
			${(top-right)}$ Positive $l$ parameter, from top to bottom $l=\left\{ 1.0,\,0.6,\,0.2\right\} $.
			${(middle-left)}$ Parameter $n_{0}$ of the component state $\left|\alpha,\xi\right\rangle $
			for $l<0$ and ${(middle-right)}$ $l>0$. ${(bottom-left)}$ Optimal squeezing angle, $\theta_{\mathsf{opt}}$ 
			for $l<0$ and ${(bottom-right)}$ $l>0$. The dashed black line in all plots corresponds the non-entangled 
			case ($l=0$), studied in a previous work \cite{douglas14}. We have used $n_{\mathsf{in}}=2.0$ and  
			$\theta = \theta_{\mathsf{opt}}$ in the plots for $H$ and $n_{0}$.}
	\end{figure}
	\par\end{center}
	
\end{document}